\documentclass[amsmath,amssymb,pra,aps,showpacs,twocolumn,10pt,superscriptaddress,tightenlines]{revtex4-2}
\usepackage[colorlinks=true,citecolor=blue,linkcolor=red,urlcolor=blue]{hyperref}
\usepackage{graphicx}
\usepackage{dcolumn}
\usepackage{bm}
\usepackage[utf8]{inputenc}
\usepackage[T1]{fontenc}
\usepackage{lmodern}
\usepackage{cleveref}
\usepackage{color}
\usepackage{amsmath}
\usepackage{amssymb}
\usepackage{amsfonts}
\usepackage{latexsym}
\usepackage{multirow}
\usepackage{booktabs}
\usepackage{comment}

\crefname{equation}{}{}

\begin{document}

\title{Quantum Random Features: A Spectral Framework for Quantum Machine Learning}

\author{Akitada Sakurai}\email{Akitada.Sakurai@oist.jp}
\affiliation{Okinawa Institute of Science and Technology Graduate University, Onna-son, Okinawa 904-0495, Japan}

\author{Aoi Hayashi}
\affiliation{Okinawa Institute of Science and Technology Graduate University, Onna-son, Okinawa 904-0495, Japan}
\affiliation{School of Multidisciplinary Science, Department of Informatics, SOKENDAI (the Graduate University for Advanced Studies), 2-1-2 Hitotsubashi, Chiyoda-ku, Tokyo 101-8430, Japan}

\author{William John Munro}
\affiliation{Okinawa Institute of Science and Technology Graduate University, Onna-son, Okinawa 904-0495, Japan}

\author{Kae Nemoto}
\affiliation{Okinawa Institute of Science and Technology Graduate University, Onna-son, Okinawa 904-0495, Japan}

\date{\today}
\begin{abstract}
\textbf{Abstract.}
Quantum machine learning (QML) models often require deep, parameterized circuits to capture complex frequency components, limiting their scalability and near-term implementation. We introduce \textit{Quantum Random Features} (QRF) and \textit{Quantum Dynamical Random Features} (QDRF), lightweight quantum reservoir models inspired by classical random Fourier features (RFF) that generate high-dimensional spectral representations without variational optimization. Using $Z$-rotation encoding combined with random permutations or Hamiltonian dynamics, these models achieve $N_f$-dimensional feature maps at preprocessing cost $O(\log(N_f))$. Spectral analysis shows that QRF and QDRF reproduce the behavior of RFF, while simulations on Fashion-MNIST reach up to 89.3\% accuracy— matching or surpassing classical baselines with scalable qubit requirements. By linking spectral theory with experimentally feasible quantum dynamics, this work provides a compact and hardware-compatible route to scalable quantum learning.
\end{abstract}

\maketitle
\section{Introduction}
Quantum machine learning (QML)~\cite{ wilson2019,Mitarai2018,Noori2020,Schuld2021, Wu2021,David2024,Gujju2024,bowles2024,sahebi2025,Sweke2025} offers a potential route to enhanced representations of high-dimensional data, but practical progress depends on architectures that are simultaneously expressive, efficient, and compatible with near-term hardware. Many existing approaches rely on parameterized quantum circuits (PQC) or quantum circuit learning (QCL) models whose outputs can be expressed as Fourier series in the encoded data~\cite{Mitarai2018, Schuld2021, Wu2021, Caro2021,Casas2023,Shin2023,sahebi2025,Sweke2025,Heimann2025}. Although increasing circuit depth expands the accessible frequency spectrum and can improve expressivity, it also introduces excessive parameters, unstable optimization, and hardware overhead~\cite{Schuld2021,okumura2025}.

Recent theoretical work has extended this spectral viewpoint to fixed-parameter models such as quantum reservoir computing (QRC)~\cite{Fujii2017,Nakajim2019,Chen2020,Angelatos2021,Domingo2021,Suzuki2022,Bravo2022} and quantum extreme learning machines (QELM)~\cite{Ghosh2019,Akitada2022,Innocenti2023,LoMonaco2024,Xiong2025,DeLorenzis2025,Sakurai2025}, which leverage naturally evolving dynamics or fixed unitaries combined with classical post-processing. While such models embed a rich set of frequency components, they lack fine control over the spectral composition, making it difficult to ensure that the relevant frequencies for a given task are covered without resorting to deep or complex circuits.

Classical Random Fourier features (RFF)~\cite{Rahimi2007} provide a scalable and theoretically grounded framework for approximating shift-invariant kernels by projecting inputs into high-dimensional, frequency-rich spaces using fixed random transformations. Inspired by the spectral correspondence between RFF and fixed-circuit QML, we introduce Quantum Random Features (QRF) and Quantum Dynamical Random Features (QDRF): layered quantum reservoir architectures that deliberately reproduce RFF-like spectral structures using experimentally feasible operations. Our design encodes data via single-qubit $R_z$ rotations interleaved with either fixed random permutations or Hamiltonian dynamics—implemented through an Ising-type interaction—to scramble phases and enrich frequency coverage. This yields $N_f$-dimensional feature maps with preprocessing cost $O(\log(N_f))$, providing exponential compression relative to classical kernel-based quantum models while remaining compatible with noisy intermediate-scale quantum (NISQ) hardware.

We show that QRF and QDRF recover the key spectral statistics of classical RFF while scaling efficiently with qubit number. On the Fashion-MNIST~\cite{Xiao2017} benchmark, the models achieve up to 89.3\% accuracy, comparable to or surpassing classical baselines under similar resources. Together, these results establish a scalable, hardware-compatible approach for constructing expressive QML architectures grounded in spectral theory.

\section{Models}
For the problem setup, we define the dataset as $\mathbb{D}_n = \{(\boldsymbol{x}_1, y_1), \cdots, (\boldsymbol{x}_n, y_n)\}$, where each input $\boldsymbol{x}_{i} \in \mathbb{R}^d$ is a $d$-dimensional vector (for image classification, this corresponds to a flattened image representation). The output $y_i$ denotes the target variable, representing either an observed value in regression tasks or a label in classification tasks. Here we focus on the multi-class image classification setting, with $c$ denoting the number of classes. Each $y_i$ is a $c$-dimensional one-hot vector whose nonzero entry marks the correct class.

\subsection{Random Fourier Features}

\begin{figure}[thb]
\centering
\includegraphics[width=0.5\textwidth]{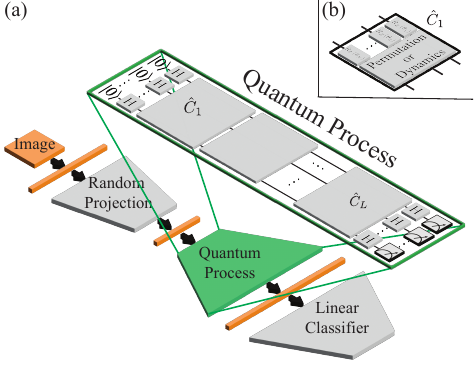}
 \caption{Schematic diagram of the QRF model. (a) the architecture consists of three key parameters: $d$ (input dimension), $N$ (number of qubits), and $L$ (number of encoding layers). (b) the circuit sequentially encodes classical data through $R_z$ rotations and applies layerwise permutations or dynamical evolutions to generate a high-dimensional quantum feature map.
 }
\label{Model}
\end{figure}

Classical Random Fourier Features (RFF) provide a scalable approach for approximating shift-invariant kernel functions by mapping inputs into high-dimensional, frequency-rich feature spaces through random linear projections (see Appendix~\ref{app:RFF}). Specifically, RFF approximate a shift-invariant kernel $k(\boldsymbol{x},\boldsymbol{x}')$ via the transformation $\varphi_{\mathrm{RFF}}(\boldsymbol{x}) = \cos(W_{\mathrm{RFF}} \boldsymbol{x} + \boldsymbol{b}_{\mathrm{RFF}})$, where $W_{\mathrm{RFF}} \in \mathbb{R}^{N_f \times d}$ has entries drawn from $\mathcal{N}(0,\sigma^2)$ and $\boldsymbol{b}_{\mathrm{RFF}} \in \mathbb{R}^{N_f}$ is sampled uniformly from $[0,2\pi)$. The resulting features are then used in a linear classifier trained with cross-entropy loss and optimized using the AdaGrad algorithm~\cite{goodfellow2016}.  In our simulation, we set learning late as 0.05, the mini-batch size as 32 unless otherwise stated and training epoch as 100.

In this work, we adopt a slightly modified version of RFF compared to the original formulation. In particular, we reinterpret RFF as a two-layer neural network whose hidden-layer weights are fixed and randomly initialized, providing a more transparent connection to neural feature learning. This perspective establishes a natural bridge between classical kernel methods and quantum circuit architectures and forms the conceptual basis for our Quantum Random Features (QRF) framework.

\subsection{Quantum Random Features (QRF)}
Because an $N$-qubit system can provide $2^N$ outputs, namely the probability distribution of bitstrings, it has the potential to generate $N_f = 2^N$ feaqture maps using only $N$ physical resources. However, a direct quantum implementation of this mapping would require computing $2^N$ rotation angles for $N$ qubits, making it impractical for large-scale applications.

To overcome this scaling bottleneck, we design a layered quantum data-encoding circuit that reproduces the spectral characteristics of RFF without explicitly constructing the full $2^N$ weight matrix. Each of the $L$ layers consists of:
\begin{enumerate}
    \item \textbf{Z-rotation encoding} — data are embedded via single-qubit $R_z(\theta)$ gates, where the angles $\theta$ are computed from a  random matrix $W \in \mathbb{R}^{(NL) \times d}$ with entries drawn from $\mathcal{N}(0, \sigma^2/NL)$, together with a bias vector $\boldsymbol{b}$ sampled uniformly from $[0, \beta)$.
    \item \textbf{Spectral scrambling} — realized through a fixed random permutation matrix $P_\pi$ (for QRF) or by Hamiltonian evolution under an Ising-type model (for QDRF introduced in the next section), both serving to randomize phases and enrich the accessible frequency spectrum.
\end{enumerate}
Measuring in the computational basis yields a probability distribution whose $N_f = 2^N$ components form the quantum feature map. As discussed in Appendix~\ref{PhaseShiftB}, we set $\beta = \pi\sqrt{3/NL}$ to prevent probability condensation and to maintain close correspondence between QRF/QDRF and the RFF formulation.

\begin{figure*}[t]
\centering
\includegraphics[width=1.0\textwidth]{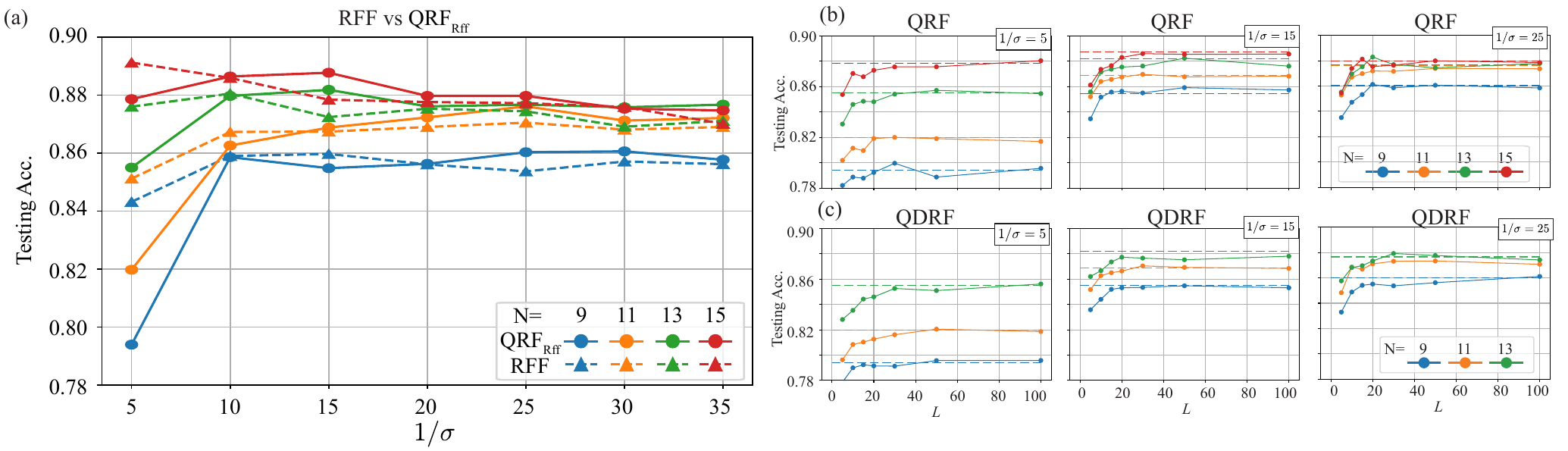}
 \caption{ 
 Test performance of (a) RFF-like Quantum Random Features ($\mathrm{QRF}_\mathrm{Rff}$), (b) Quantum Random Features (QRF), and (c) Quantum Dynamical Random Features (QDRF). In (a) comparison between classical RFF and the extreme quantum model as a function of $1/\sigma$ for $N = 10\sim 13$. In panel (b), QRF performance converges toward that of the extreme case as the number of layers $L$ increases, with $L \approx 20$–$30$ sufficient for all tested qubit counts and kernel widths $\sigma$. This result shows that QRF attains exponential feature expressivity with only linear preprocessing cost. In panel (c), QDRF employs Ising-model dynamics to achieve spectral scrambling and reaches accuracy comparable to the extreme case, demonstrating that natural Hamiltonian evolution offers an effective and experimentally compatible alternative to random permutations. In this simulation, we set $g/J=1.0$, $Jt=3.5$ and $\alpha=1.5$, where $t$ is a duration time per layers.
}
\label{CQRFvsRFFvsQRFvsQDRF}
\end{figure*}

The random frequencies of QRF are represented by $\boldsymbol{\Theta} = BW \cdot \boldsymbol{x}$, which serves as the counterpart to $W_{\mathrm{RFF}}$ in the RFF model. Here, the matrix $B$ is constructed from the diagonal elements of the Pauli operators $\sigma^z_{i \in \{1,\cdots, N\}}$ combined with the permutation operator $\hat{P}_{\pi}$, whose formal definition is provided in Appendix~\ref{QRF_Gaussian}. Therefore, as the number of layers $L$ increases, the composite transformation $BW$ asymptotically approaches the statistical properties of $W_{\mathrm{RFF}}$. Empirically, we find that for $L \approx 20$, $BW$ already provides a close approximation to $W_{QEC}$ (see Appendix~\ref{QRF_Gaussian}).The overall preprocessing cost of the QRF model scales as $O(NLd)$; since typically $L \ll 2^N$ for large qubit counts, this represents a substantial reduction compared with the classical RFF formulation.

\subsection{Quantum Dynamical Random Features (QDRF)}

Implementing arbitrary permutation matrices on current noisy intermediate-scale quantum (NISQ) hardware is costly due to the increased circuit depth and fidelity demands of multi-qubit operations. To enhance experimental feasibility, we replace $\hat{P}_\pi$ in QRF with naturally realizable unitary evolutions generated by simple, high-fidelity Hamiltonians. Specifically, we employ the long-range transverse-field Ising model:
\[
\hat{H} = \sum_{i>j} \frac{J}{|i-j|^\alpha} \sigma_z^i \sigma_z^j + g \sum_i \sigma_x^i,
\]
where $J$ and $g$ denote the coupling and field strengths, and $\alpha$ controls the spatial decay of interactions. For numerical simulations, we set $\alpha = 1.5$, a regime consistent with recent trapped-ion experiments~\cite{Porras2004, Zhang2017, Zhang2017Ising}. This Hamiltonian formulation enables QDRF to preserve the spectral scrambling capability of QRF while remaining directly compatible with accessible NISQ hardware.

\section{Numerical Simulations}
As a benchmark for testing Expressivity and Accuracy: RFF, QRF, and QDRF, we use the Fashion-MNIST dataset~\cite{Xiao2017}, which contains 60{,}000 grayscale images for training and 10{,}000 for testing. Each image is an intensity map with pixel values originally ranging from 0 to 255, representing 10 clothing categories. In our simulations, all pixel values are normalized to the range $[0, 1]$ prior to encoding.

As discussed in the previous section, increasing the number of layers causes the random frequency components of QRF to converge toward those of the classical RFF. We therefore first compare RFF with a RFF-like Quantum Random Features ($\mathrm{QRF}_\mathrm{Rff}$) in which the quantum model is explicitly constructed to reproduce the same random frequencies as RFF. This model is the direct quantum implementation, as discussed in the previous section. This cases's quantum state prior to measurement is given by
\[
|\phi(\boldsymbol{\Theta})\rangle = \hat{H}^{\otimes N} |\boldsymbol{\Theta} \rangle,
\]
where $|\boldsymbol{\Theta}\rangle = \frac{1}{\sqrt{2^N}}\sum_{l=0}^{2^N-1} e^{i\theta_{l+1}} |l\rangle,$
and the phase angles $\boldsymbol{\Theta} = (\theta_{1},\cdots,\theta_{2^N})^\top$ are defined as $\boldsymbol{\Theta} = W_{\mathrm{RFF}}\boldsymbol{x} + \boldsymbol{b}$. Fig.~\ref{CQRFvsRFFvsQRFvsQDRF} (a) compares the test accuracy of RFF and $\mathrm{QRF}_\mathrm{Rff}$ across different qubit numbers and kernel bandwidths $\sigma$. The results show that the quantum model achieves performance comparable to RFF; in particular, with 16 qubits and $1/\sigma = 10$, it attains a test accuracy of 89.3\%, surpassing previously reported quantum machine learning baselines. We emphasize, however, that this computational advantage does not yet manifest in $\mathrm{QRF}_\mathrm{Rff}$, since the exponential cost of constructing $W_{\mathrm{RFF}}$ remains.

Now, we compare the results of QRF with $\mathrm{QRF}_\mathrm{Rff}$. Results across different numbers of qubits and layers are shown in Fig.~\ref{CQRFvsRFFvsQRFvsQDRF} (b). As the number of layers $L$ increases, the test accuracy of QRF gradually converges toward that of $\mathrm{QRF}_\mathrm{Rff}$. For any number of qubits and $\sigma$, $L \approx 20$–$30$ is sufficient to achieve comparable performance. This demonstrates that we have successfully reduced the computational cost of QRF, for example, with a compression ratio $\gamma = NL/2^N = 0.06$ for $N = 12$ and $L = 20$. Importantly, the required number of layers remains nearly constant as the number of qubits increases. Thus, while the number of features grows exponentially with $N$, the preprocessing cost scales only linearly, resulting in an exponentially improved computational efficiency. Here, we compare our result with other models that also use a reservoir or fixed middle network similart to ours. Our results are equal to or greater than those of these models~\cite{lau2024,Lorenzis2025,Sakurai2025}. Here, we emphasize that our model does not utilize well-tuned pre-processing techniques, such as PCA and Autoencoder, unlike other models.

In addition to the benchmark of the Fahion-MNIST, we have applied the QRF to the CIFAR-10 dataset~\cite{Krizhevsky09} which consistes of 60,000 RGB imgage with 10 classes. We used 50,000 images and 10,000 images for training and testing. The QRF reaches about 53\% testing accuracy with $N=14$, $1/\sigma=25$ and $L=50$, and we set the learning rate 0.005.

To demonstrate the necessity of entanglement in the unitary, we also tested a non-entangling permutation matrix consisting solely of single-qubit operations. As shown in Fig.~\ref{Th_Nu_In}, its weight matrix does not show the Gaussian property. Consequently, 
the maximum test accuracy for 10 to 14 qubits was consistently below 75\%. These results show that random single-qubit rotations fail to reproduce the desired performance, highlighting the essential role of entangling capability in the unitary operator.

Finally, we examine the results of the Quantum Dynamical Random Features (QDRF) model, in which Ising-type Hamiltonian dynamics replace the fixed permutation matrix. Fig.~\ref{CQRFvsRFFvsQRFvsQDRF} (c) shows that QDRF performs comparably to $\mathrm{QRF}_\mathrm{Rff}$, demonstrating that naturally evolving Hamiltonian dynamics can effectively serve as a substitute for permutation-based spectral scrambling. The role of the permutation (or dynamical) operator is to generate a pseudo-random matrix $B$ by reordering computational basis states, thereby enhancing spectral diversity and inducing multi-qubit entanglement.

\subsection{Robustness Under Finite Sampling}

\begin{figure}[htb]
\centering
\includegraphics[width=0.4\textwidth]{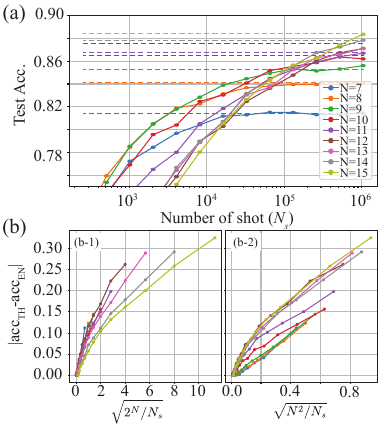}
 \caption{
 Number of shots required. (a) shows the test accuracy obtained from the theoretical probability distribution and from empirical distributions computed with different shot numbers ranging from $2^8$ to $2^{20}$. The dashed lines represent the reference accuracy based on the theoretical distribution. (b) plots the absolute difference between theoretical and empirical accuracies as a function of the scaling variables $\sqrt{2^N/N_s}$ and $\sqrt{N^2/N_s}$. Here, $\mathrm{acc}_\mathrm{TH}$ and $\mathrm{acc}_\mathrm{EN}$ represent the testing accuracy using theoretical and empirical probability distributions, respectively. The parameters are set to $1/\sigma = 15$ and $L = 30$.
 }
\label{Sampling}
\end{figure}

We have not yet accounted for the sampling (finite-shot) cost required to obtain the probability distribution of the quantum system in our model. In this section, we numerically evaluate the minimum number of measurement shots required to estimate this distribution. Specifically, we compute the empirical distribution through numerical simulations, train the linear classifier using this empirical data, and then test its performance with the same empirical distributions. Fig.~\ref{Sampling} (a) shows the relationship between the number of shots and test accuracy for systems with 8 to 14 qubits, together with the accuracy achieved using the exact theoretical distribution. As expected, the accuracy approaches the theoretical limit as the number of shots increases. Notably, even with relatively few shots, test accuracy remains well above random guessing. This demonstrates that the linear classifier is robust to sampling noise, allowing the model to maintain useful performance even with incomplete probability estimates.

To investigate scaling with system size, we examined the absolute difference between test accuracies obtained from theoretical and empirical distributions under rescaled axes. We considered two rescaling factors, $\sqrt{2^N/N_s}$, which is widely believed in the QML community~\cite{Thanasilp2024aa,Xiong2025,kairon2025,sannia2025,saem2025}, and $\sqrt{N^2/N_s}$, where $N_s$ is the number of shots. The results, shown in Fig.~\ref{Sampling}(b-1) and Fig.~\ref{Sampling}(b-2), reveal that the deviation does not scale as $O(2^N)$; instead, it follows a trend closer to $O(N^2)$, albeit with slight deviations. These findings indicate that in our model, the required number of shots does not grow exponentially with the number of qubits. We confirmed the same scaling behavior on the MNIST dataset of handwritten digits. Interestingly, while both datasets show the same scaling, their convergence rates differ: MNIST converges more rapidly than Fashion-MNIST, likely due to differences in dataset complexity and classification difficulty.

The observed $N^2$ scaling results reveal two distinct regimes. The first corresponds to systems with 7 to 9 qubits, where the Hilbert space dimension is smaller than the number of pixels in the original image (the input dimension). The second encompasses 11 to 15 qubits, where the Hilbert space dimension exceeds the input dimension. This bifurcation arises from whether the quantum feature map projects the input into a higher-dimensional space—that is, whether a high-dimensional embedding is realized. These findings emphasize the importance of high-dimensional mappings in quantum random feature models, indicating a qualitative transition in behavior once the quantum feature space surpasses the input dimension.

\section{Connections to Existing QML Frameworks}

This section explores the relationship between our proposed models—Quantum Random Features (QRF) and Quantum Dynamical Random Features (QDRF)—and other quantum machine learning (QML) approaches. As noted in the introduction, the encoding strategy in QRF/QDRF, which maps classical data into quantum states via layered circuits, resembles frameworks used in Quantum Circuit Learning (QCL)~\cite{Gil2020, PerezSalinas2020, Schuld2021} and Quantum Extreme Learning Machines (QELM)~\cite{Xiong2025}. QCL involves optimizing the quantum circuit for a specific task, while QELM uses a fixed quantum circuit followed by a classical learning step. Our models are conceptually more closely aligned with QELM, since the quantum component is fixed and only the classical classifier is trained. 

Our model can also be viewed as a variant of optimization-based models~\cite{Gil2020, Schuld2021,Casas2023,Shin2023,Heimann2025}, where the parameterized circuit replaces the permutation matrix and no final linear classifier is used. Both our model and theirs apply a classical transformation (e.g., an affine map or a random projection) before encoding. Conceptually, the models are similar, but key differences arise in how spectral coverage and scalability are achieved.

All models address function approximation via Fourier analysis of layered quantum encodings~\cite{Gil2020, Schuld2021,Casas2023,Shin2023,Heimann2025}. The layered structure enables control over the frequency spectrum, allowing these models to approximate target functions with sufficient depth and width. In optimization-based approaches, the circuit parameters control the Fourier coefficients, while in non-optimized models such as QRF, this responsibility shifts to the downstream classical part.

Despite theoretical progress, applying these models to practical tasks (e.g., image classification) remains challenging. The core challenge lies in ensuring adequate coverage of the relevant spectral components, which is difficult without prior knowledge. This typically requires increasing the number of qubits or circuit depth, both of which are experimentally demanding for high-dimensional data such as images.
Our QRF model addresses this issue by extending classical random feature methods to the quantum domain. It leverages the well-established utility of large random networks, both in theory and in practice. We provide analytical insights alongside empirical results demonstrating that QRF is practical and effective for real-world tasks.

We now compare QRF with Quantum Extreme Reservoir Computing (QERC)~\cite{Akitada2022}, which employs natural quantum dynamics (i.e., Hamiltonian evolution) for image classification. The quantum circuit in QERC can be viewed as a special case of QRF. However, QRF alone does not reach the performance of QERC. For example, QRF typically requires around 20 layers, while QERC achieves high accuracy with just two, due to its use of PCA for dimensionality reduction. When PCA is applied to QRF inputs with nine qubits ($N = 9$) and $L = 2$, we achieve a test accuracy of approximately 84\%, which is significantly higher than the approximately 79\% achieved without PCA.

PCA~\cite{Mackiewicz1993aa} has a computational cost of $O(D^3+d^2D)$ due to matrix diagonalization. However, using approximation methods such as randomized SVD~\cite{Liberty2007aa,Woolfe2008aa,rokhlin2009} and reducing the number of components to $2N$ lowers this to $O(DNd)$. In contrast, QRF's preprocessing cost is $O(dNL)$, which is advantageous for large $D$. While training is more computationally intensive for QRF, inference is significantly more efficient than for QERC. Thus, QRF is preferable when ample training resources are available, whereas QERC is better suited to resource-limited or real-time inference settings.

In classical models, the input dimension and feature count are independently tunable, though high performance may require their alignment. QERC's encoding and output dimensions, however, depend on the number of qubits, with output scaling exponentially. This makes dimensionality reduction via PCA essential but fundamentally limited—especially for high-resolution data such as medical images. QRF offers greater flexibility: its circuit depth and output dimension can be tuned independently, making it more adaptable and scalable for complex, high-dimensional inputs.

\section{Conclusion and Outlook}

In this work, we proposed Quantum Random Features (QRF) and Quantum Dynamical Random Features (QDRF), quantum generalizations of classical random feature methods that deliver expressive, scalable, and hardware-compatible mappings for quantum machine learning (QML). Through theoretical analysis and numerical experiments, we showed that QRF and QDRF efficiently reproduce the behavior of Random Fourier Features (RFF) while remaining compatible with current NISQ devices.

Our results reveal that the spectral structure underlying RFF can be naturally realized within shallow, layered quantum circuits. By adjusting the number of encoding layers, QRF achieves exponential expressive capacity at linear preprocessing cost, whereas QDRF exploits Hamiltonian dynamics to enable effective spectral scrambling and direct hardware implementation. Compared with parameterized circuit models, QRF and QDRF eliminate the need for variational optimization—avoiding barren plateaus and instability—and enable transparent control over the frequency content of encoded data. Relative to reservoir-based approaches such as QERC, our framework provides a clearer link between spectral theory and physical realizability, unifying expressivity, scalability, and experimental accessibility.

Beyond image classification, this framework offers a foundation for quantum-enhanced learning in domains such as time-series forecasting, generative modeling, and reinforcement learning. Future directions include task-specific spectral engineering and formal characterization of expressivity, sample complexity, and generalization performance, which will deepen our understanding of how spectral structure governs quantum feature mappings.

Overall, QRF and QDRF represent versatile and theoretically grounded building blocks for practical QML architectures. By bridging spectral theory with experimentally feasible quantum dynamics, they provide a unified and scalable route toward near-term quantum learning models.

\acknowledgements{

This work was partly supported by the MEXT Quantum Leap Flagship Program (MEXT Q-LEAP) under Grant No.~JPMXS0118069605, the COI-NEXT program under Grant No.~JPMJPF2221,the JSPS KAKENHI Grant No.~21H04880, and JSPS KAKENHI under Grant No. 25K21306
}

\appendix
\section{From Random Fourier Features to Quantum Random Features}\label{app:RFF}

In this appendix, we review how \textit{Random Fourier Features} (RFF) arise in the context of kernel methods, and then derive their quantum analoge, referred to as \textit{Quantum Random Features} (QRF).

\subsection{Random Fourier Features and Bochner’s theorem}

Consider a shift-invariant kernel,
\begin{equation}
    k(x,y) = k(x-y).
\end{equation}
By Bochner’s theorem, such kernels can be expressed as the Fourier transform of a positive measure:
\begin{equation}
    k(x-y) = \int_{\mathbb{R}^d} e^{i \omega \cdot (x-y)} \, p(\omega) \, d\omega,
\end{equation}
where $p(\omega)$ is the spectral density of the kernel. Approximating this integral by Monte Carlo sampling with $D$ random frequencies $\{\omega_j\}_{j=1}^D$ gives the random feature mapping:
\begin{equation}
    \phi(x) = \sqrt{\tfrac{2}{D}} \Big[ \cos(\omega_1 \cdot x + b_1), \dots, \cos(\omega_D \cdot x + b_D) \Big]^\top,
\end{equation}
where $\omega_j \sim p(\omega)$ and $b_j \sim \mathrm{Unif}[0,2\pi]$.  
The kernel can then be approximated as
\begin{equation}
    k(x,y) \approx \phi(x)^\top \phi(y).
\end{equation}
This formulation provides a computationally efficient means of approximating nonlinear kernel functions using linear feature mappings.

\subsection{Neural network interpretation}

An alternative perspective is to view RFF as a two-layer neural network:
\begin{itemize}
    \item The first layer applies a random affine transformation, $W_{\text{RFF}} x + b$,  
    \item The second layer applies a cosine nonlinearity.
\end{itemize}
Explicitly,
\begin{equation}
    \phi(x) = \sqrt{\tfrac{2}{D}} \, \cos\!\left(W_{\text{RFF}} x + b\right),
\end{equation}
where $W_{\text{RFF}} \in \mathbb{R}^{D \times d}$ has rows drawn from $p(\omega)$.  
Thus, the RFF representation can be interpreted as a random two-layer neural network with cosine activations, providing a conceptual bridge between kernel methods and neural networks.

\subsection{Quantum analog: layered encodings}

To construct a quantum analog, we replace the cosine nonlinearity with interference patterns arising from quantum circuit dynamics.
Consider a system of $N$ qubits with input vector $x \in \mathbb{R}^d$.  
We define the encoding unitary as
\begin{equation}
    \hat{U}(x) = \hat{P}_\pi \, \hat{R}_z(Wx + b),
\end{equation}
where $\hat{R}_z(\theta)$ denotes single-qubit $Z$ rotations by angles determined by the affine map $Wx + b$, and $\hat{H}^{\otimes N}$ applies Hadamard gates to all qubits.  

Measuring in the computational basis produces feature probabilities:
\begin{equation}
    p(l \mid x) = \left| \langle l | \hat{H}^{\otimes N} \hat{U}(x) |0\rangle^{\otimes N} \right|^2,
\end{equation}
which serve as the quantum random features.  
Here, quantum interference plays the role of the nonlinear activation, providing a direct physical mechanism analogous to the cosine nonlinearity in RFF.

By stacking $L$ layers of the form
\begin{equation}
    \hat{U}_\ell(x) = \hat{P}_{\pi} \, \hat{R}_z(W_\ell x + b_\ell), \qquad \ell = 1, \dots, L,
\end{equation}
we obtain the overall feature map:
\begin{equation}
    |\psi(x)\rangle = \hat{U}_L(x) \cdots \hat{U}_1(x) |0\rangle^{\otimes N}.
\end{equation}
Increasing $L$ enriches the spectral coverage of the encoded features, in direct analogy to increasing the number of random neurons in the classical RFF model. This multi-layered construction formally defines the Quantum Random Features (QRF) model.

\section{Approximating Classical Random Features with Quantum Random Features}\label{QRF_Gaussian}

In this section, we verify that Quantum Random Features (QRF) approximate the behavior of the $\mathrm{QRF}_\mathrm{Rff}$ discussed in the main text. Specifically, we analyze the relationship between the phase angles of the complex coefficients in the computational basis states generated by the QRF circuit—comprising $Z$-axis rotation gates and permutation operators—and those of the corresponding state of the $\mathrm{QRF}_\mathrm{Rff}$. For simplicity, we assume $\beta = 0$ throughout this discussion.

\subsection{Effective large linear transformation}

Since the QRF model employs only phase encoding and permutation, the state of the system just before the final Hadamard gates closely resembles that of the $\mathrm{QRF}_\mathrm{Rff}$. Therefore, the QRF state can be represented as
\begin{equation}
|{\boldsymbol{\Theta}_{PQ}}\rangle = \frac{1}{\sqrt{2^N}} 
\left(e^{i\theta_1}, \cdots, e^{i\theta_{2^N}}\right)^\top,
\end{equation}
where $\boldsymbol{\Theta}_{PQ} = \left(\theta_1, \cdots, \theta_N\right)$ denotes the phase vector generated by the phase encoding and permutation. We now derive the relationship between $\boldsymbol{\Theta}_{PQ}$ and the input parameters through the following linear transformation:
\begin{equation}
\boldsymbol{\Theta}_{PQ} =
B \left(W \cdot \boldsymbol{x} + \boldsymbol{b}\right),
\end{equation}
where $B$ is defined as
\begin{equation}
B  = 
\left[
\begin{matrix}
(P_{\pi})^L \Sigma_N  & (P_{\pi})^{L-1} \Sigma_N  & \cdots & (P_{\pi}) \Sigma_N
\end{matrix}
\right].
\label{eq:Bmatrix}
\end{equation}
Here, the matrix $B$ consists of the permutation matrix $P_\pi$, and $\Sigma_N$ is defined by
\begin{equation}
\Sigma_N = \left[
\begin{matrix}
\mathrm{diag}(\sigma_1^z) & \mathrm{diag}(\sigma_2^z) & \cdots & \mathrm{diag}(\sigma_N^z)
\end{matrix}
\right],
\end{equation}
where $\mathrm{diag}(\sigma_i^z)$ denotes the diagonal elements of $\sigma_i^z$ in vector form. Because of the random permutation, the matrix $B$ statistically behaves as a random matrix.

\subsection{Gaussianity of $V=B W$}
In the $\mathrm{QRF}_\mathrm{Rff}$, the transformation from the input $\boldsymbol{x}$ to the angles $\boldsymbol{\Theta}$ is given by $\boldsymbol{\Theta} = W_{QEC} \cdot \boldsymbol{x}$, where $W_{QEC}$ is sampled from the normal distribution $\mathcal{N}(0, \sigma^2)$. In contrast, for the QRF, the corresponding transformation is $\boldsymbol{\Theta} = BW \cdot \boldsymbol{x}$. Thus, it is crucial to determine whether $V = BW$ can also be regarded as sampled from the same normal distribution. We therefore examine the statistical properties of the matrix $V = BW$ in terms of its mean and variance. The $(i,j)$ element of $V$ is given by
\begin{equation}
V_{ij} = \sum_{k=1}^{NL} B_{ik} W_{kj}.
\end{equation}
Since $B_{ik} \in \{-1,1\}$, $B_{ik} W_{kj}$ represents a random sign flip of $W_{kj}$. Due to the symmetry of the normal distribution, it follows the same distribution $\mathcal{N}(0, \sigma^2/NL)$. Furthermore, because these terms are independent across different indices $k$, the central limit theorem applies to their sum over $NL$ independent normal variables. Consequently, $V_{ij}$ is normally distributed with variance $NL \times \sigma^2 / NL = \sigma^2$, implying that $V = BW$ exhibits the same statistical distribution as the original normal distribution.

\subsection{Independence of row vectors of $V=B W$}

We have observed that the elements $V_{ij}$ appear to be sampled from a normal distribution $\mathcal{N}(0, \sigma^2)$. However, this observation alone is insufficient; we must also examine the independence of the column vectors. Since the row vectors of the weight matrix correspond to the frequency components of the input, it is desirable to minimize correlations between them. To simplify the discussion, we assume that the matrix $B \in \{-1,1\}^{2^N \times NL}$ is drawn from a Bernoulli distribution with equal probabilities. The two row vectors $\boldsymbol{V}_{(i, \cdot)}$ and $\boldsymbol{V}_{(j, \cdot)}$ are given by $\boldsymbol{V}_{(i, k)} = \sum_l B_{i, l} W_{lk}$ and $\boldsymbol{V}_{(j, k)} = \sum_l B_{j, l} W_{lk}$. Their inner product is then
\begin{equation}
\begin{split}
\langle \boldsymbol{V}_{i, \cdot}, \boldsymbol{V}_{j, \cdot} \rangle = \sum_{k, l, m} B_{i, l} B_{j, m} W_{lk} W_{mk}.
\end{split}
\end{equation}

The expectation of this inner product is
\begin{equation}
\mathbb{E}\!\left[ \langle \boldsymbol{V}_{i, \cdot}, \boldsymbol{V}_{j, \cdot} \rangle \right] = d \sigma^2 \delta_{i, j}.
\end{equation}

Hence, the expected self-similarity (for $i = j$) is $d \sigma^2$, while the expected cross-similarity (for $i \neq j$) vanishes. Although the expectation value is zero for $i \neq j$, the variance may still be significant. Evaluating this variance gives
\begin{equation}
\mathbb{V}\!\left[ \langle \boldsymbol{V}_{i, \cdot}, \boldsymbol{V}_{j \neq i, \cdot} \rangle \right] = d \sigma^4 \left( 1 + \frac{1}{NL} + \frac{d}{NL} \right).
\end{equation}
Here, the asymptotic limit of the above expression, obtained by taking $N$ and $L$ to infinity, coincides with that of the Gaussian distribution.

The ratio between the standard deviation of the similarity for $i \neq j$ and the expected self-similarity for $i = j$ is given by
\begin{equation}
\begin{split}
\gamma_{V} &= \frac{\sqrt{\mathbb{V}\!\left[ \langle \boldsymbol{V}_{i, \cdot}, \boldsymbol{V}_{j, \cdot} \rangle \right]}}{\mathbb{E}\!\left[ \langle \boldsymbol{V}_{i, \cdot}, \boldsymbol{V}_{i, \cdot} \rangle \right]} = \sigma^2 \sqrt{\frac{1}{d} + \frac{1}{NL} + \frac{1}{dNL}}.
\end{split}
\end{equation}
Since $\sigma < 1$ and $d$, $N$, and $L$ are integers, if $\sigma < (1/3)^{1/4}$, then $\gamma_{V}$ is less than one for any $d$, $N$, and $L$. Moreover, as $d$, $N$, and $L$ increase, correlations between different row vectors of $\boldsymbol{V}$ diminish and asymptotically concentrate around zero.

\begin{figure}[b]
\centering
\includegraphics[width=0.48\textwidth]{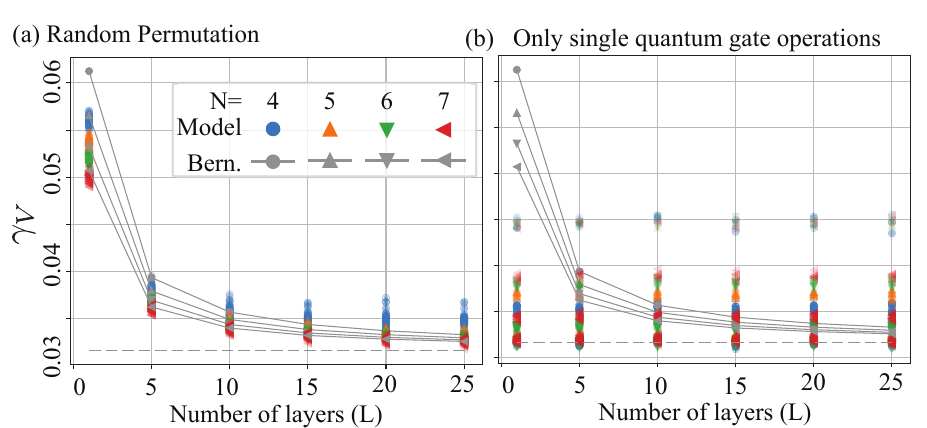}
 \caption{
 Comparison of theoretical and numerical results for row vector correlations. The solid gray line represents the theoretical values calculated from the Bernoulli distribution, while the colored dots show numerical results obtained using permutation matrices. For each trial, 100 random pairs of indices $(i, j)$ are selected to compute $\langle \boldsymbol{V}_{i, \cdot}, \boldsymbol{V}_{j, \cdot} \rangle$. The gray dashed line indicates the asymptotic value $\sigma^2 / \sqrt{d}$. Each data point is averaged over $10^4$ realizations to estimate the standard deviation $\mathbb{V}\!\left[ \langle \boldsymbol{V}_{i, \cdot}, \boldsymbol{V}_{j, \cdot} \rangle \right]$. Different marker shapes correspond to different numbers of qubits. In this comparison, we set $\sigma = 1/10$.
 }
\label{Th_Nu_In}
\end{figure}

Previously, we theoretically showed that the row vector correlations are extremely small, assuming that the matrix $B$ follows a Bernoulli distribution. However, in the actual model, as defined in Eq.~\ref{eq:Bmatrix}, $B$ is constructed using randomly sampled permutation matrices. These permutation-based constructions effectively cause $B$ to approximate the statistical behavior of a Bernoulli-distributed matrix. To investigate this correspondence, we performed numerical simulations in which $\gamma_{V}$ was computed over $10^4$ trials for systems with 4 to 8 qubits. In each trial, both Gaussian random matrices and random permutation matrices were independently generated. For the latter, we evaluated the inner product between the first and second row vectors to estimate correlation behavior.

The results are shown in Fig.~\ref{Th_Nu_In}(a). The solid gray line represents the theoretical calculations, while the colored dots correspond to the results obtained using permutation matrices.  Note that for an $N$-qubit system, the total number of distinct index pairs $(i, j)$ with $i \neq j$ is $2^N(2^N - 1)/2$; however, in the plot, we display results for 100 randomly selected pairs. The gray dashed line indicates the asymptotic value $\sigma^2 / \sqrt{d}$. Different marker shapes denote different numbers of qubits. As shown in the figure, the results obtained with permutation matrices exhibit small deviations from those predicted by the Bernoulli model. Nevertheless, as the number of layers $L$ increases, the permutation-matrix results converge toward the asymptotic line at $\sigma^2 / \sqrt{d}$. Therefore, the row vector correlations remain sufficiently small for practical applications, even when permutation matrices are used.

When considering a quantum machine learning model, an important question is whether the model requires two-qubit (entangling) gates. If the entire process can be implemented using only single-qubit gates, it can be efficiently simulated on a classical computer. In this model, generating a random permutation requires at least one entangling gate. However, if the permutation can be represented using only single-qubit operations, we can also examine this restricted scenario. As in the previous example, we generated such limited permutations randomly and computed $\gamma_V$, as shown in Fig.~\ref{Th_Nu_In}(b). The resulting values differ significantly from those obtained under the Bernoulli assumption. Interestingly, the results remain nearly constant across all qubit numbers, indicating that the correlation value does not decrease as the number of layers increases. This occurs because a purely single-qubit permutation matrix reverts to its original configuration after two applications. From this, we conclude that such matrices cannot sufficiently scramble the frequency components. Therefore, the inclusion of entangling gates is essential for reproducing the statistical properties characteristic of the Bernoulli-distributed case.

\subsection{The Role of the Phase Shift $b$}
\label{PhaseShiftB}

Next, we discuss the role of the phase-shift parameter $\boldsymbol{b}$. Both $\boldsymbol{b}$ and $\boldsymbol{b}_{QEC}$ help suppress the trivial localization of the quantum system’s probability distribution. To illustrate this, we consider the the $\mathrm{QRF}_\mathrm{Rff}$ as an example. First, consider the probability distribution of the $\mathrm{QRF}_\mathrm{Rff}$ when $\boldsymbol{b}_{QEC} = 0$. In most machine learning models, the input $\boldsymbol{x}$ is normalized to have a magnitude of 1. Assuming that $\boldsymbol{x}$ is sampled from a uniform distribution, the expected value of the phase angle $\boldsymbol{\Theta}$ is given by $\mathbb{E}[\theta] = 0$ with variance $\mathbb{V}[\theta] = {d\sigma^2}/{3}$. Here, we assume independence between $\boldsymbol{x}$ and $W_{QEC}$. Similar to RFF, we take $\sigma < 1$. Consequently, when $d < 3 / \sigma^2$, the phase angles $\theta$ are concentrated near zero.

We now analyze the corresponding probability distribution in the regime $d < 3 / \sigma^2$. The distribution is obtained by applying the Hadamard gate to the state $|{\boldsymbol{\Theta}}\rangle$ and then measuring it in the computational basis. The probability amplitude $A_{00\cdots 0}$ for the computational basis state $|0 \cdots 0\rangle$ is given by $A_{00\cdots 0} = \frac{1}{\sqrt{2^N}} \sum_{l=1}^{2^N} e^{i\theta_l},$
where we have used $H_{1l} = 1 / \sqrt{2^N}$, representing the elements of the Hadamard matrix. Since the phase angles are localized around zero, we can approximate $e^{i\theta} \approx 1$ for $\theta \ll 1$, leading to $A_{00\cdots 0} \approx 1$. As a result, the computational basis state $|00 \cdots 0\rangle$ is measured with almost unit probability. As the threshold $d$ scales inversely with $\sigma^2$, this localization effect persists even for moderately large $d$.

In QRF, the phase-shift vector $\boldsymbol{b}$ is scrambled by the matrix $B$, and the total phase shift $\boldsymbol{b}_{Q}$ is given by $\boldsymbol{b}_{Q} = B \cdot \boldsymbol{b}$. By the central limit theorem, the elements of $\boldsymbol{b}_{Q}$ approximately follow a Gaussian distribution. Here again, we assume that $B$ is sampled from a Bernoulli distribution. The standard deviation of $\boldsymbol{b}_{Q}$ is then given by
\begin{equation}
\sqrt{\mathbb{V}[\boldsymbol{b}_{Q}]} = \beta \sqrt{\frac{NL}{3}}.
\end{equation}
In RFF and CQRF, the phase shift parameter is uniformly distributed over $[0, 2\pi)$, implying a standard deviation of $\pi$. Therefore, by equating the standard deviations, we obtain
\begin{equation}
\beta = \pi \sqrt{\frac{3}{NL}}.
\end{equation}

\end{document}